# Interpreting New Data on Large Scale Bulk Flows

**Richard Watkins** [1]

*Physics Department, University of Michigan*
*Ann Arbor, MI 48109-1120*

**Hume A. Feldman**[2]

*Physics Department, Princeton University*
*Princeton, NJ 08544*

## ABSTRACT

We study the implications of a recent estimate of the bulk flow of a sample of galaxies containing supernovae type Ia by Riess, Press, and Kirshner. We find that their results are quite consistent with power spectra from several currently popular models of structure formation, but that the sample is as yet too sparse to put significant constraints on the power spectrum. We compare this new result with that of Lauer and Postman, with which there is apparent disagreement. We find that for the power spectra we consider, the difference in window functions between the two samples used for the measurements results in a low level of expected correlation between the estimated bulk flows. We calculate a $\chi^2$ for the two measurements taken together and find that their lack of agreement tends to disfavor spectra with excessive power on large scales, but not at a level sufficient to rule them out. A sample consisting of other SN type Ia's found in the Asiago catalog is used to study how the sensitivity of the method used by RPK will improve with increasing sample size. We conclude that the local group motion should be able to be determined with a sample of $\sim 100$ SN Ia light curve shapes.

*Subject headings*: cosmology: distance scales – cosmology: large scale structure of the universe – cosmology: observation – cosmology: theory – galaxies: kinematics and dynamics – galaxies: statistics

[1]Address as of 1 Sep. 1995: Dept of Physics & Astronomy, Dartmouth College, Hanover, NH 03755

[2]feldman@tatania.princeton.edu

## 1. Introduction

Recently, Riess, Press & Kirshner (1995a,1995b) (hereafter RPK) have developed a distance measure applicable for high redshift galaxies ($\sim 10,000$ km/s) using type Ia supernova (SN Ia) light curve shapes that can achieve accuracies estimated to be about 5% of the redshift. This is a significant improvement over previous high redshift methods, most notably that based on brightest cluster galaxies (BCG) used by Lauer and Postman (1994) (hereafter LP), which has an estimated accuracy of 17% of the redshift. Using their distance measure, RPK have reported the bulk flow of a sample of 13 distant galaxies obtained primarily from the Calán/Tololo supernova search (Hamuy 1993, Maza *et al.* 1994, see also RPK 1995a,b and references therein) in which SN Ia light curves have been observed; a sample with a median redshift of $\approx 5500$ km/sec. They report their result as being inconsistent at a high confidence level ($\sim 99\%$) with the result of Lauer and Postman, who measured the bulk flow of a complete, volume limited sample of 119 Abell clusters with a comparable depth. This result is especially intriguing in light of the fact that several groups have shown that the LP bulk flow is inconsistent at the $\gtrsim 95\%$ level with the power spectra of most currently popular models (Feldman & Watkins 1994, Strauss *et al.* 1994, Tegmark *et al.* 1994; see also Jaffe & Kaiser 1994). As we shall show, the RPK result is quite consistent with all the power spectra we have considered.

In this *Letter* we explore the implications of this new measurement for our knowledge of the power spectrum on large scales. We calculate $\chi^2$ statistics for the RPK result by itself and the RPK and LP results taken together for a variety of power spectra. We also calculate a measure of the expected correlation between results from the LP and RPK samples. We find that the lack of agreement between the two measurements is not surprising in the context of currently popular power spectra; for these spectra we expect both the RPK and LP measurements to be dominated by noise and incomplete cancelation of smaller scale flows and thus to be nearly uncorrelated. Spectra with greatly enhanced large scale power, as favored by the LP measurement, predict stronger agreement between the two bulk flow vectors and are disfavored by the inclusion of the RPK result. Finally, we discuss the expected sensitivity of the RPK SN Ia sample for determining the underlying bulk flow of the volume in which the sample is embedded, and examine how this sensitivity will improve as the sample size increases.

## 2. Analysis

The analysis in this letter follows closely our previous analysis of the LP survey (Feldman & Watkins 1994, see also Kaiser 1988). Here we give an overview of our methods and refer the reader to our previous paper for details. We also expand our analysis to include the comparison of two data sets sampling the same peculiar velocity field.

Given a bulk flow vector $U_i$, the covariance matrix for the estimated bulk flow of a sample of galaxies is the sum of two statistically independent parts,

$$R_{ij} \equiv \langle U_i U_j \rangle = R_{ij}^{(v)} + R_{ij}^{(\varepsilon)} \ , \tag{1}$$

the first part arising from the sampling of the underlying velocity field and the second arising due to the noise in the distance estimates. The velocity part of the covariance matrix,

$$R_{ij}^{(v)} = \int \frac{d^3k}{(2\pi)^3}\, \mathcal{W}_{ij}^2(\vec{k})\ P_v(\vec{k}) \ , \tag{2}$$

is the convolution of the velocity power spectrum with $\mathcal{W}_{ij}^2(\vec{k}) = W_{il}(\vec{k})\ W_{jm}(\vec{k})\ \hat{k}_l\ \hat{k}_m$ , where $W_{ij}$ is the tensor window function of the sample, given in Fourier space by

$$W_{ij}(\vec{k}) = A_{im}^{-1} \sum_n \frac{\hat{r}_{n,m}\ \hat{r}_{n,j}}{\sigma_n^2 + \sigma_*^2}\ e^{i\vec{k}\cdot\vec{r}_n} \ . \tag{3}$$

In these equations and the equations to follow, repeated indices denote implicit sums. Here $\sigma_*$ is the dispersion in the line–of–sight velocity due to random velocities (which we take to be a constant for the sample), $\sigma_n$ is the estimated uncertainty in the line–of–sight peculiar velocity, $\hat{r}_{n,j}$ is the $j$th component of the unit vector of the $n$th galaxy, and

$$A_{ij} = \sum_n \frac{\hat{r}_{n,i}\ \hat{r}_{n,j}}{\sigma_n^2 + \sigma_*^2}\ . \tag{4}$$

The uncertainty $\sigma_n$ is typically proportional to the redshift of the galaxy; for the RPK sample we use the distance errors reported for each galaxy (Riess *et al.* 1995b). The choice of $\sigma_*$ affects mainly the noise term in the covariance. We shall adopt a value of 400 km/s, consistent with the value of 382 km/s obtained by dividing the recent pairwise velocity difference estimate from the CfA2+SSRS2 redshift surveys (Marzke *et al.* 1995) by $\sqrt{2}$. The velocity power spectrum is given by $P_v(k) \equiv \langle |v(\vec{k})|^2 \rangle = \frac{H^2 a^2}{k^2} P(k)$, where $P(k)$ is the density power spectrum. The noise term in the covariance matrix is simply $R_{ij}^{(\epsilon)} = A_{ij}^{-1}$.

We define a $\chi^2$ statistic for the three degrees of freedom of the measured bulk flow vector $\vec{V}$ to be given by $\chi_V^2 \equiv V_i\ R_{ij}^{-1}\ V_j$, where $V_i$ is the $i$th component of $\vec{V}$. $R_{ij}$ can also be used to calculate an expectation value for the magnitude of the bulk flow denoted by $\Lambda$, a convenient number with which to compare different spectra and catalogs. We also calculate the expectation values for the velocity ($\Lambda^{(v)}$) and noise ($\Lambda^{(\epsilon)}$) parts of the covariance matrix separately.

In addition to applying our method to the RPK sample of 13 SN Ia's, we also wish to study how the sensitivity of the RPK method will increase as the sample size becomes larger. In order to do this, we have obtained a sample of 61 SN Ia's taken primarily from the Asiago catalog of recent supernovae (including the 13 used by RPK). This sample should be typical of SN Ia samples gathered in the future. We have applied our analysis to subsets of this sample of varying sizes as well as to larger mock samples designed to have roughly the same distribution in redshift.

In order to study the likelihood that a given power spectrum could have produced both the LP and RPK results, we construct the 6-dimensional vector $\vec{U}^T = (\vec{U}^{LP}; \vec{U}^{RPK})$. Using a similar analysis to that described above, we calculate a covariance matrix $R^T_{ij} \equiv \langle U^T_i U^T_j \rangle$ and a corresponding $\chi^2$ for 6 degrees of freedom given by $\chi^2_T \equiv U^T_i \ (R^T)^{-1}_{ij} \ U^T_j$. Additionally, we get an idea of how much correlation we expect between $\vec{U}^{LP}$ and $\vec{U}^{RPK}$ for a given power spectrum by calculating the normalized expectation value for their dot–product,

$$\mathcal{C} = \frac{\langle U^{LP}_i U^{RPK}_i \rangle}{(\langle U^{LP}_l U^{LP}_l \rangle \langle U^{RPK}_m U^{RPK}_m \rangle)^{\frac{1}{2}}} \ , \tag{5}$$

which should be close to 1 for highly correlated vectors, zero for vectors that are completely uncorrelated, and $-1$ if there is a high degree of anti-correlation.

We consider power spectra from the IRAS–QDOT survey (Feldman, Kaiser & Peacock 1994), the BBKS standard CDM ($\sigma_8 = 1$, $\Omega h = 0.5$) model (Bardeen *et al.* 1986), a CDM spectrum ($\mathrm{CDM_{LS}}$) which fits IRAS–QDOT well at intermediate scales, but has lots of power on large scales ($\Omega h = 0.075$,$\sigma_8 = 0.9$), and a PIB generated power spectrum ($\Omega = 0.1$, $\lambda = 0.9$, Peebles 1994). We have corrected the IRAS–QDOT spectrum for the redshift distortion pointed out first by Kaiser (1987). In addition, since it has no information for $k < 0.025 \ h\mathrm{Mpc}^{-1}$ corresponding to scales $\gtrsim 250 \ h^{-1}$ Mpc, we have extrapolated to the COBE point so that it is well defined for all scales of interest (see Fig 1a). For comparison, we have also included results for the limiting case of a spectrum ($\mathrm{PS}_{k\to 0}$) which has *all* of its power at $k \to 0$, normalized to give maximum likelihood for the LP data set (Feldman, Jaffe, Kaiser, & Watkins 1995). It should be noted that both this spectrum and $\mathrm{CDM_{LS}}$ conflict strongly with the COBE measurements.

We have seen that the velocity part of the covariance matrix is the integral over $k$ space of the product of the squared tensor window function and the velocity power spectrum [see Eq. (2)]. Study of the squared window function of a survey is therefore useful for determining which scales contribute to the bulk flow estimate and to what degree. A sparse sample will tend to pick up contributions from small scales through incomplete cancelation. This is exhibited by the window function not falling to zero outside of the central peak, but rather approaching a constant. The height of this "plateau" decreases roughly as $N$, the number of sample objects. For sparse samples, the components of the tensor window function vary rapidly with the angle and magnitude of $\vec{k}$. These small scale variations tend to smooth out on averaging over angles. In what follows, all window functions are assumed to be averaged over angles and are plotted as a function of $|\vec{k}|$.

In Fig. 1b we show the trace of the squared tensor window function for the RPK sample. For comparison, we include the trace of the squared tensor window function for the LP survey. From this figure it is clear that, except on the largest scales, the RPK and LP samples probe the power spectrum in very different ways; indeed, if one were to look at the full three dimensional window function, the overlap between the two window functions would appear to be even smaller. This implies that while both vectors will have similar contributions from the very largest scales, contributions from smaller scales will in general not be correlated.

In Fig. 1c we show the trace of the squared tensor window function for mock surveys with the same distribution in redshift as the RPK sample. The window functions are shown for samples of varying sizes, averaged over 20 realizations for each size. Averaging over many realizations tends to smooth out features associated with specific placements of sample objects, making the underlying "plateau" more prominent.

In table 1, we show $\chi^2$ for the RPK result, the LP result, and both results taken together using a variety of power spectra. We also include the measure $\mathcal{C}$ of the expected correlation [Eq. (6)]. Note that the RPK result is quite consistent with all the power spectra we have considered. Also note that the LP values of $\chi^2$ are somewhat larger than we reported in our previous work (Feldman & Watkins 1994) due to the correction of a small error in our calculation.

If there were no overlap between the RPK and LP window function, then the resulting bulk flow vectors would be expected to be uncorrelated and $\chi^2_T$ would be the sum of $\chi^2_{RPK}$ and $\chi^2_{LP}$. Window function overlap gives cross–terms which tend to favor agreement between the two vectors; *i.e.* if the window functions are similar then the vectors should be too. Here, since the RPK and LP bulk flow vectors point in almost opposite directions, overlap will increase $\chi^2_T$ so that the probability of measuring both vectors decreases. Power spectra with lots of power on large scales, where the overlap is greatest, will be more strongly disfavored due to the higher expectation for correlation between the two results. As we see from table 1, this effect is most important for the $\mathrm{CDM}_{\mathrm{LS}}$ and $\mathrm{PS}_{k\to 0}$ spectra. For the other spectra, the large value of $\chi^2_T$ can be attributed almost entirely to the large value of $\chi^2_{LP}$; indeed, the inclusion of the RPK data increases the likelihood for the IRAS–QDOT and CDM spectra.

In Fig. 2, we present results of our analysis using mock catalogs of varying sizes as discussed above. For these catalogs, we give the noise-free expectation value $\Lambda^{(v)}$ for each of our power spectra, as well as the noise expectation value. For each of the sample sizes, we created a number of random catalogs and averaged over the results. The variance over different random catalogs was of order 10% and has been neglected.

As the number of objects in the sample increases, the noise contribution to the expectation value decreases like $N$. For small $N$, the velocity contribution to the expectation value will also decrease like $N$, due to the fact that with a small number of objects the largest part of this term is coming from the incomplete cancelation of small scale flows (see Fig. 1c). As the volume becomes well sampled, however, the velocity expectation approaches a constant which represents the velocity of the volume as a whole. This effect is clearly seen in Fig. 2. Since the sensitivity is highest in the galactic $z$ direction, we have also included in Fig. 2 a plot showing the expectations for this component alone.

## 3. Discussion

Comparison of the results of the RPK and LP studies assumes that they are measuring the same quantity. However, an examination of Fig. 1b and Table 1 shows that this is not necessarily the case. The RPK and LP bulk flow vectors contain significant contributions due to noise and incomplete cancelation of small scale flows. Both of these contributions depend on the details of the survey and of the power spectrum and would not be expected to correlate across different samples. As we have discussed above, the effect of the disagreement between the RPK and LP results is to disfavor models with large amounts of power on large scales, although not at a level that provides significant constraints. However, the RPK results by themselves and taken together with LP suggest that we may be able to describe the large scale velocity field without a major revision to theories of large scale structure formation.

Clearly, if RPK type measurements are to effectively constrain the power spectrum, many more objects will be needed. If we assume that the power spectrum on large scales is not too far different from conventional spectra, *e.g.* that of CDM or IRAS–QDOT, then from Fig. 2 we can estimate that the signal to noise should become $\approx 1$ when the number of SN Ia's in the sample is of order 100. When the number of objects reaches 200, one should have a fairly precise value for the bulk flow of the sample. Given the greater sensitivity in the $z$ direction, it is likely that the SN Ia estimate for the $z$ component of the bulk flow could be reasonably accurate with just 60 or so objects. Between the Calán/Tololo survey and the CfA collection of supernovae, the RPK sample size should reach 40 by the end of 1995. However, the ending of the Calán/Tololo search effort makes the prospects dim for significantly increasing this number in the near future.

In contrast, for an LP type survey, with typical distance errors of approximately 15%, a sample size of the order of 300 is needed to get a a signal to noise of about one for conventional power spectra, or about 200 data points for the $z$ component $V_z$. Of course, if the actual bulk velocity is larger than the expectation values we have calculated, than it will show in a sparser survey; indeed, LP may have already detected a large $z$ component for the bulk flow. One disadvantage of using clusters is that the number *density* of objects in the survey cannot be increased (it is a fairly complete sample). Increasing the sample volume to reduce the noise should also result in a smaller bulk flow signal, and thus the signal to noise may not increase as much as one might expect. Significantly increasing the signal to noise for studies of this type may require decreasing the error in the distance measurements to $\lesssim 10\%$.

**Acknowlegement:** We thank Andrew Jaffe, Tod Lauer, and Michael Strauss for useful comments, and Adam Riess for comments and helpful assistance in obtaining the data for both the RPK sample and the Asiago catalog of recent supernovae. HAF was supported in part by the National Science Foundation and by NASA grants NAG5–1310 and NAG5–2412. RW was supported in part by NASA grant NAG5–2869.

---

**Table 1: $\chi^2$ for RPK, LP and Total**

| Spectrum | $\chi^2_{LP}$ | $P(\chi^2 > \chi^2_{LP})$ | $\chi^2_{RPK}$ | $P(\chi^2 > \chi^2_{RPK})$ | $\chi^2_T$ | $P(\chi^2 > \chi^2_T)$ | $\mathcal{C}$ |
|---|---|---|---|---|---|---|---|
| QDOT | 10.40 | 0.015 | 2.82 | 0.420 | 14.11 | 0.028 | 0.08 |
| CDM | 10.41 | 0.015 | 2.58 | 0.461 | 13.81 | 0.032 | 0.07 |
| $\mathrm{PS}_{(k\to 0)}$ | 2.39 | 0.495 | 0.70 | 0.873 | 14.04 | 0.029 | 0.70 |
| $\mathrm{CDM}_{\mathrm{LS}}$ | 5.52 | 0.137 | 1.28 | 0.734 | 10.84 | 0.093 | 0.35 |
| PIB | 11.33 | 0.010 | 3.43 | 0.323 | 16.81 | 0.010 | 0.11 |

The $\chi^2$ and Probability for the LP and RPK surveys and the total quantities. Note that the quantity $P(\chi^2 > \chi^2_T)$ is calculated for six degrees of freedom whereas $P(\chi^2 > \chi^2_{LP})$ and $P(\chi^2 > \chi^2_{RPK})$ are calculated for three degrees of freedom.

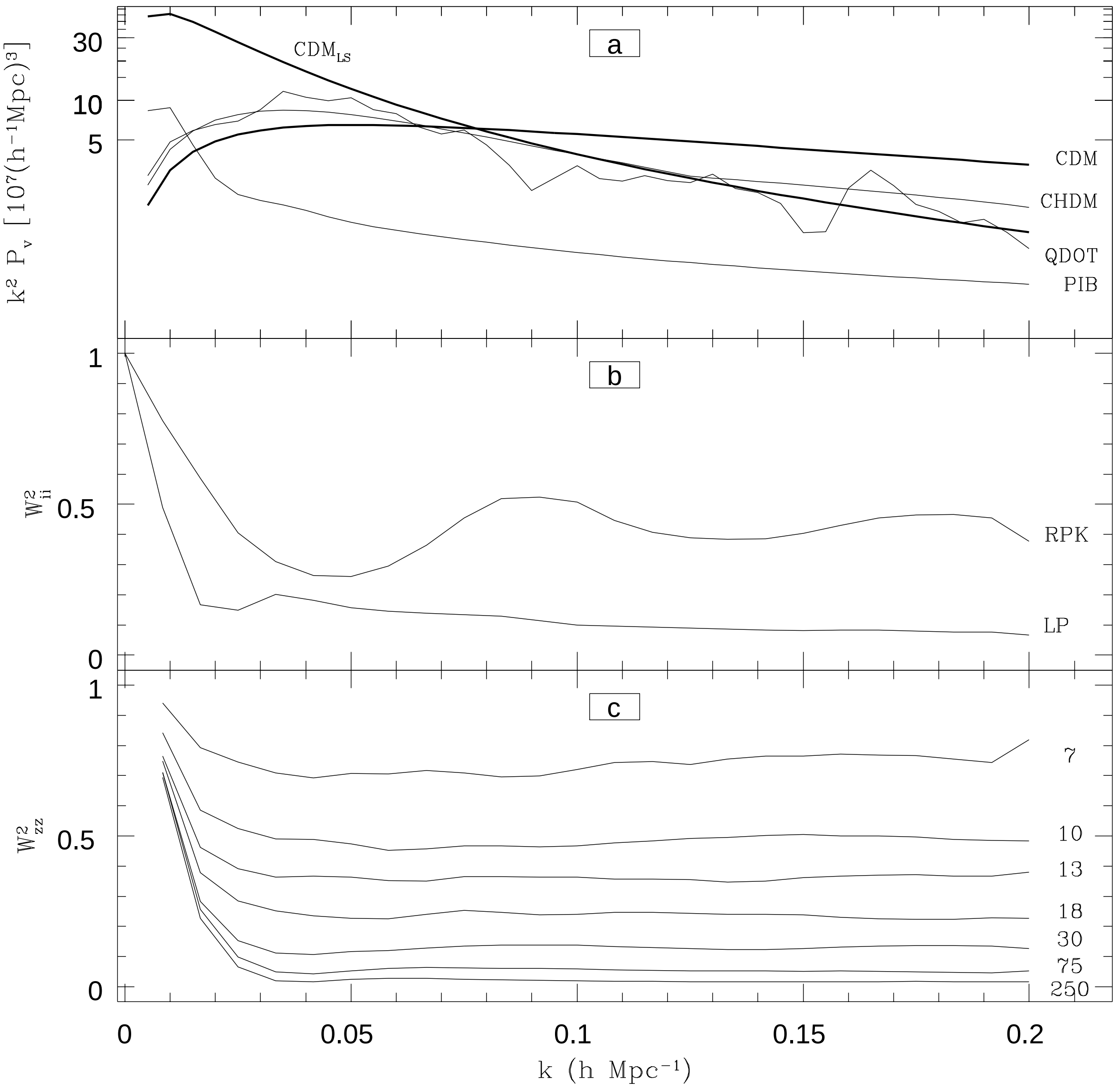

Fig. 1.— [a] The redshift corrected power spectra used in the analysis. [b] The trace of the squared tensor window functions for the RPK survey, as well as the LP ones. [c] The window functions for SN Ia like surveys of different sizes. The contributions from large $k$ fall as the number of data points increase.

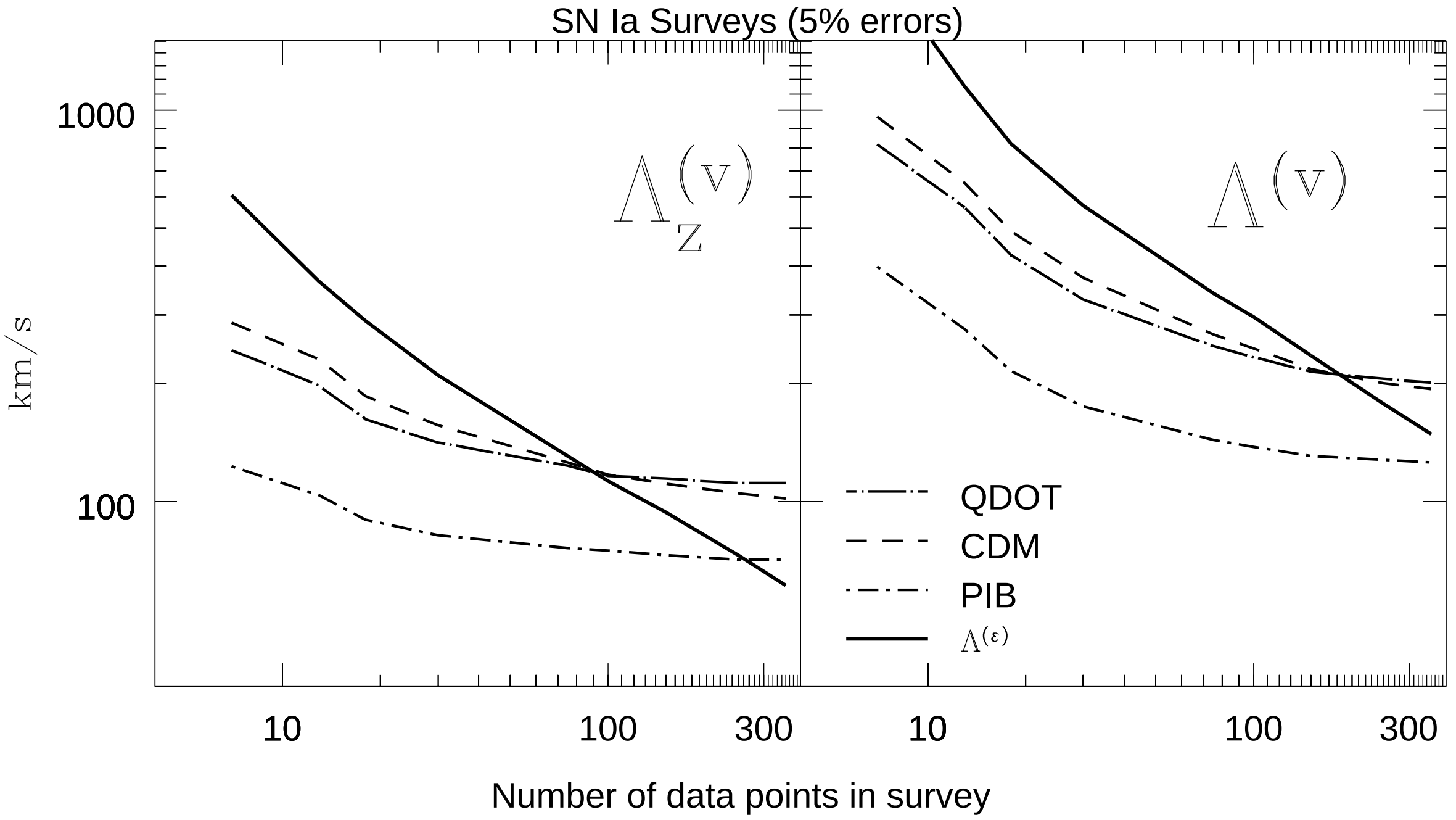

Fig. 2.— The noise-free expectation values for the $z$ component of $\Lambda^{(v)}$ and its magnitude for the power spectra considered as a function of the size of the survey. We also show the expected magnitude of the noise, $\Lambda^{(\varepsilon)}$, which falls with the number of data points.